\documentclass[%
    aps,
    prd,
    reprint,
    10pt,
    superscriptaddress,
    amsfonts,
    amssymb,
    amsmath,
    preprintnumbers,
    noshowpacs,
    tightenlines,
    floats,
    notitlepage,
    final,
    letterpaper,
    oneside,
    twocolumn,
    nofootinbib,
    longbibliography,
    ]{revtex4-2}
\pdfoutput=1
\usepackage[utf8]{inputenc}
\usepackage[T1]{fontenc}
\usepackage[british]{babel}
\usepackage[british,cleanlook,printdayon]{isodate}
\usepackage[Symbolsmallscale]{upgreek}
\usepackage{mathtools}
\usepackage{isomath}
\usepackage[protrusion,tracking,kerning,spacing,final,babel]{microtype}
\usepackage{physics}
\usepackage[dvipsnames]{xcolor}
\usepackage[final]{hyperref}
\usepackage[normalem]{ulem}
\hypersetup{%
    colorlinks=true,
    citecolor=MidnightBlue,
    linkcolor=MidnightBlue,
    urlcolor=MidnightBlue
    }%
\usepackage{cleveref}
\usepackage{tikz}
\usetikzlibrary{shapes.geometric,arrows}
\tikzstyle{flow} = [rectangle, rounded corners, minimum width=2cm, minimum height=0.5cm,text centered, draw=black, fill=blue!30]
\tikzstyle{arrow} = [thick,->,>=stealth]

\newcommand{\rme}{\mathrm{e}}
\newcommand{\rmi}{\mathrm{i}}

\begin{document}

\begin{flushright}
    KCL-PH-TH-2022-42
\end{flushright}

\title{Gravitational-wave event rates as a new probe for dark matter microphysics}
\date{\today}

\author{Markus R.~Mosbech}
\email{markus.mosbech@sydney.edu.au}
\affiliation{School of Physics, The University of Sydney, Camperdown NSW 2006, Australia\\
ARC Centre of Excellence for Dark Matter Particle Physics\\
Sydney Consortium for Particle Physics and Cosmology
}

\author{Alexander C.~Jenkins}
\email{alex.jenkins@ucl.ac.uk}
\affiliation{Department of Physics and Astronomy, University College London, London WC1E 6BT, United Kingdom}

\author{Sownak Bose}
\email{sownak.bose@durham.ac.uk}
\affiliation{Institute for Computational Cosmology, Department of Physics, Durham University, Durham DH1 3LE, United Kingdom}

\author{Celine Boehm}
\email{celine.boehm@sydney.edu.au}
\affiliation{School of Physics, The University of Sydney, Camperdown NSW 2006, Australia\\
ARC Centre of Excellence for Dark Matter Particle Physics\\
Sydney Consortium for Particle Physics and Cosmology
}

\author{Mairi Sakellariadou}
\email{mairi.sakellariadou@kcl.ac.uk}
\affiliation{Theoretical Particle Physics and Cosmology Group,  Physics Department, King's College London, University of London, Strand, London WC2R 2LS, United Kingdom}

\author{Yvonne Y.~Y.~Wong}
\email{yvonne.y.wong@unsw.edu.au}
\affiliation{School of Physics, The University of New South Wales, Sydney NSW 2052, Australia\\
Sydney Consortium for Particle Physics and Cosmology}

\begin{abstract}
    We show that gravitational waves have the potential to unravel the microphysical properties of dark matter due to the dependence of the binary black hole merger rate on cosmic structure formation, which is itself highly dependent on the dark matter scenario.
    In particular, we demonstrate that suppression of small-scale structure --- such as that caused by interacting, warm, or fuzzy dark matter --- leads to a significant reduction in the rate of binary black hole mergers at redshifts $z\gtrsim5$.
    This shows that future gravitational-wave observations will provide a new probe of physics beyond the $\Lambda$CDM model.
\end{abstract}

\maketitle

\section{Introduction}

The standard $\Lambda$CDM model of cosmology has been posited to explain a range of observations spanning the largest observable scales to the scale of galaxies. 
Its success at explaining these data relies on two mysterious components: dark energy in the form of a cosmological constant ($\Lambda$) and dark matter (DM).
Dark matter, in particular, is key to explaining the formation and evolution of structures such as galaxies and clusters of galaxies in the late-time Universe.

Present observations on supergalactic scales are compatible with the hypothesis that the dark matter is cold --- i.e., the particles have vanishingly small thermal velocities. In this cold dark matter (CDM) model, the particles also do not have significant nongravitational interactions~\cite{Boehm:2001hm, Planck2018:1,Bull:2015stt}.
These requirements can be satisfied by particlelike dark matter of typical mass $\gtrsim$~keV, or wavelike dark matter of mass $\gtrsim 10^{-22}$~eV.
However, the key to determining the fundamental nature of dark matter lies in subgalactic scales at large redshifts.
This is because the onset of nonlinear structure formation can be very sensitive to the microphysics of the dark matter~\cite{Boehm:2000gq,Boehm:2003xr,Boehm:2004th}.
In turn, any probe that could shed light on this epoch of structure formation would also provide valuable insights into the nature of dark matter~\cite{Carucci:2015bra,Chakrabarti:2022cbu}.

\begin{figure}[t]
    \centering
    \includegraphics[width=0.48\textwidth]{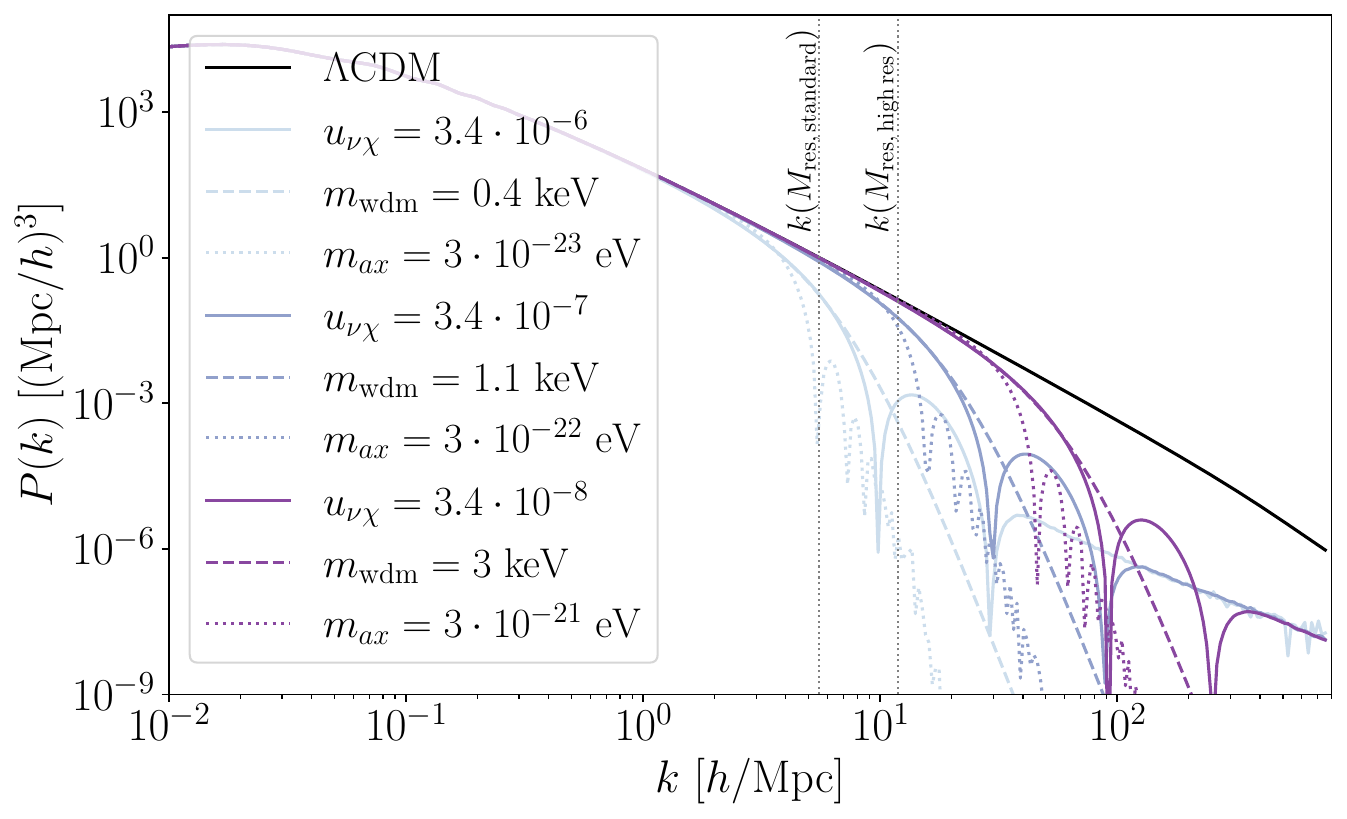}
    \caption{\label{fig:linear}
    Linear matter power spectra for several DM-neutrino interacting scenarios of varying interaction strength $u_{\nu\chi}$, along with predictions of their ``equivalent'' FDM and thermal WDM scenarios characterised by their respective particle masses, $m_{ax}$ and $m_{\rm wdm}$, tuned to give a similar small-scale suppression.
    The FDM power spectra have been computed using \textsc{AxionCAMB}~\cite{Hlozek:2014lca,axcamb}; the WDM power spectra are derived from the $\Lambda$CDM power spectrum filtered with a transfer function~\cite{Bode:2000gq}.
    Vertical lines indicate rough $k$-scales corresponding to the mass resolutions used in \textsc{galform}.}
\end{figure}

There are three classes of phenomenological particlelike and wavelike dark matter scenarios that generically predict small-scale signatures that differ from the predictions of standard CDM.
The first, warm dark matter (WDM) scenario usually assumes negligible interactions but has a small DM particle mass in the low keV range that allows these particles to free-stream out of small-scale perturbations~\cite{Dodelson:1993je,Bode:2000gq,Hansen:2001zv,Asaka:2005an,Viel:2005qj,Boyarsky:2009ix,Viel:2013fqw,Abazajian:2001nj,Boyarsky:2008xj,Dolgov:2000ew,Bose:2016}.
The second, interacting dark matter (IDM) scenario makes no strong assumption about the particle mass but endows the DM particle with non-negligible interactions (see Sec.~\ref{sec:structureformation} for details).
The third, fuzzy dark matter (FDM) scenario comprises a condensate of ultra-light DM particles of mass $\sim 10^{-22}$--$10^{-21}$~eV  whose collective behaviour is wavelike~\cite{Hui:2021tkt,Armengaud:2017nkf,Irsic:2017yje,Lidz:2018fqo}, although Lyman-$\alpha$ data implies a mass $>10^{-20}$~eV~\cite{Rogers:2020ltq}.
As shown in Fig.~\ref{fig:linear}, although the details differ between WDM, IDM, and FDM, all three scenarios predict a cutoff in the linear matter power spectrum at large wave numbers $k$, a feature that is often invoked as a possible solution to the claimed ``small-scale crisis'', a long-standing set of discrepancies where simulations predict more small-scale structures than observed~\cite{Bullock:2017xww}.

In this article we present a new observational probe of dark matter microphysics, namely the merger rate of stellar-mass binary black holes (BBHs) throughout cosmic time.\footnote{%
    We focus here on stellar-mass BBHs (and therefore Hz-band detectors such as LIGO, Virgo, Einstein Telescope, and Cosmic Explorer) because the physics underpinning their merger rates is much better understood at present than that of the supermassive BBHs probed by GW searches at lower frequencies (principally by pulsar timing arrays).}
With the recent advent of gravitational-wave astronomy, advanced interferometers like LIGO~\cite{LIGOScientific:2014pky} and Virgo~\cite{VIRGO:2014yos} have already detected dozens of BBHs~\cite{LIGOScientific:2021usb,LIGOScientific:2021djp}.
Next-generation observatories such as Einstein Telescope~\cite{Punturo:2010zz} and Cosmic Explorer~\cite{Reitze:2019iox} will be capable of detecting nearly \emph{all} stellar-mass binary black holes in the observable Universe~\cite{Regimbau:2016ike}, extending the reach of gravitational-wave astronomy out to the beginning of cosmic star formation.
The space-based interferometer LISA~\cite{LISA:2017pwj} will also provide a great deal of complementary low-frequency information about this population.
The abundances of these systems at different epochs encode the star formation, chemical enrichment, and other baryonic processes occurring within dark matter haloes.
As we demonstrate below, \emph{the BBH merger rate is highly sensitive to the suppression of small-scale structure induced by dark matter microphysics}, as this in turn inhibits star formation and reduces the BBH yield, particularly at redshifts $z\gtrsim5$.

\begin{figure}
    \begin{tikzpicture}[node distance=1.5cm]
        \node (top) {};
        \node (class) [flow, below of = top] {\textsc{class}};
        \node (galform) [flow, below right of = class] {\textsc{galform}};
        \node (gadget) [flow, below left of = class] {\textsc{gadget}};
        \node (compas) [flow, below of = galform] {\textsc{compas}};
        \node (bottom) [below of = compas] {};
        \draw [arrow] (top) -- node[anchor=west, xshift=0.25cm] {DM-$\nu$ interaction strength, $u_{\nu\chi}$} (class);
        \draw [arrow] (class) -- node[anchor=west, xshift=0.25cm] {linear DM power spectrum, $P(k)$} (galform);
        \draw [arrow] (class) -- node [] {} (gadget);
        \draw [dashed,arrow] (gadget) to[bend right] node [yshift=-0.275cm,xshift=-0.5cm] {halo mass function} (galform);
        \draw [arrow] (galform) -- node[anchor=west, xshift=0.25cm] {star formation; metallicities} (compas);
        \draw [arrow] (compas) -- node[anchor=west, xshift=0.25cm] {BBH merger rate, $\mathcal{R}_\mathrm{BBH}(z)$} (bottom);
    \end{tikzpicture}
    \caption{\label{fig:flowchart}
    An illustration of our pipeline. Linear power spectra of IDM scenarios are computed using \textsc{class}, which are then fed into \textsc{gadget} and \textsc{galform} as initial conditions. The \textsc{galform} output is cross-checked with the halo mass function from \textsc{gadget} and fed into \textsc{compas}, which computes the gravitational-wave event rate.}
\end{figure}
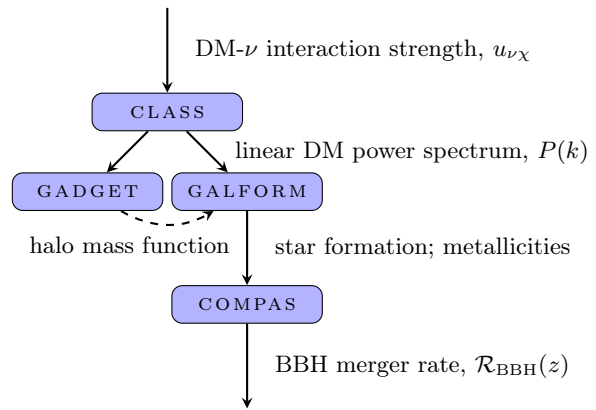

In the following, we use as a working example a DM-neutrino interacting scenario~\cite{Boehm:2000gq,Boehm:2004th,Mangano:2006mp,2010PhRvD..81d3507S,Wilkinson:2014ksa,DiValentino:2017oaw,Stadler:2019dii,Ali-Haimoud:2015pwa,nudm,HooperLucca} to demonstrate quantitatively the potential of the BBH merger rate to probe cosmologies with suppressed small-scale structure.
A brief description of this IDM scenario is given in Sec.~\ref{sec:structureformation}.
Our analysis pipeline, shown in Fig.~\ref{fig:flowchart}, is as follows.
We compute the linear matter power spectrum of an IDM cosmology using \textsc{class}~\cite{Blas:2011rf,Lesgourgues:2011rh,nudm}, which is then fed into the semianalytic model of structure formation \textsc{galform}~\cite{Cole2000,Lacey2016} to produce synthetic realisations of galactic populations, as described in Sec.~\ref{sec:galform}.  Concurrently, we perform an $N$-body simulation of the IDM cosmology using the \textsc{gadget-4} code~\cite{Gadget4,supereasy}, initialised with the same linear input, in order to cross-check the halo mass functions predicted by \textsc{galform}.  Finally, the star formation rate and stellar metallicities computed by \textsc{galform} are fed into the binary population synthesis code~\textsc{compas}~\cite{Compas,Stevenson:2017tfq,Vigna-Gomez:2018dza}, described in Sec.~\ref{sec:gravwaves},
which computes the gravitational-wave event rate.  Our results and conclusions are presented in Sec.~\ref{sec:summary}.

Note that while we have chosen to focus on a specific IDM scenario, we expect our general conclusions to remain valid for other dark matter scenarios that predict qualitatively similar suppression of small-scale power, as nonlinear evolution at $z\lesssim10$ tends to wash out their detailed features~\cite{Mosbech:2022uud}.
These include WDM and to some extent FDM as already illustrated in Fig.~\ref{fig:linear}, as well as DM interactions with itself~\cite{Carlson:1992fn,deLaix:1995vi,Spergel:1999mh,Dave:2000ar,Creasey:2016jaq,Rocha:2012jg,Kim:2016ujt,Huo:2017vef,Markevitch:2003at,Randall:2007ph,Boehm:2000gq,Boehm:2004th}, with photons~\cite{Boehm:2000gq,Boehm:2001hm,Boehm:2004th,Sigurdson:2004zp,Wilkinson:2013kia,Boehm:2014vja,Schewtschenko:2014fca,Schewtschenko:2015rno,Escudero:2015yka,McDermott:2010pa,Diacoumis:2017hff,Ali-Haimoud:2015pwa,Stadler:2018jin,CyrRacine:2012fz,Dolgov:2013una,Stadler:2018dsa}, baryons~\cite{Boehm:2000gq,Boehm:2004th,Chen:2002yh,Dvorkin:2013cea,Dolgov:2013una,CyrRacine:2012fz,Prinz:1998ua,Boddy:2018kfv,Slatyer:2018aqg,Xu:2018efh,Boddy:2018wzy},
 and dark radiation~\cite{Das:2017nub,Kaplan:2009de,Diamanti:2012tg,Buen-Abad:2015ova,Lesgourgues:2015wza,Ko:2017uyb,Escudero:2018thh,Das:2010ts,Archidiacono:2019wdp,Blennow:2012de,Cyr-Racine:2013fsa,Bose:2019}.
The FDM and baryon-IDM models may need more detailed analysis to confirm that their impact on baryons on subgalactic scales does not fundamentally differ from CDM.

\section{Interacting dark matter}\label{sec:structureformation}

We consider the case in which the dark matter scatters elastically with (massive) neutrinos, via a constant, velocity-independent cross section $\sigma_0$ as described in~\cite{nudm}.
This kind of interaction can be realised in some particle physics theories beyond the Standard Model~\cite{Boehm:2003hm,Boehm:2003xr,Boehm:2006mi,Olivares-DelCampo:2017feq}.
From the phenomenological perspective, however, the cosmological dynamics of this interaction can be captured by a single parameter, the interaction strength, normally parameterised in terms of the dimensionless quantity~\cite{Wilkinson:2013kia,DiValentino:2017oaw,Stadler:2019dii,nudm}
    \begin{equation}
        u_{\nu\chi}\equiv\frac{\sigma_0}{\sigma_\mathrm{Th}}{\qty(\frac{m_\chi}{100\,\mathrm{GeV/c^2}})}^{-1},
    \end{equation}
where $m_\chi$ is the DM mass, $\sigma_0$ the interaction cross section, and $\sigma_\mathrm{Th}\approx6.65\times10^{-29}\,\mathrm{m}^2$ is the Thomson scattering cross section.
The case of DM-massive neutrino interaction was previously implemented in the Einstein-Boltzmann solver \textsc{class}~\cite{Blas:2011rf,Lesgourgues:2011rh} in~\cite{nudm}, which we use in this work to generate the linear matter power spectra for our IDM scenarios.

As noted earlier, the interaction suppresses the matter power spectrum below a certain scale, similar to the predictions of WDM scenarios.
As collisions occur between dark matter and neutrinos, the former is prevented from collapsing to form structures, leading to a suppressed linear matter power through collisional damping.
This type of damping is physically distinct from free-streaming in WDM scenarios, where the large initial velocities result in the WDM particles escaping from initial overdensities.
It is also distinct from the damping seen in FDM scenarios, which arises from quantum pressure hindering gravitational collapse.
In general, the stronger the coupling, the later the DM decouples from the neutrinos and hence the larger the scales affected (see Fig.~\ref{fig:linear}).
Observe also that IDM scenarios predict in addition prominent acoustic oscillations at higher wave numbers.
The origin of these acoustic oscillations is analogous to the baryon acoustic oscillations (BAO) from the coupling of photons and baryons at early times, and the scale at which the oscillations appear is again governed by the dark matter and neutrino decoupling times.

Cosmic microwave background (CMB) anisotropy and BAO measurements currently constrain the DM-neutrino interaction strength to $u_{\nu\chi}\lesssim10^{-4}$ (95\% confidence)~\cite{nudm}.
Including Lyman-$\alpha$ forest measurements strengthens the bound to $u_{\nu\chi}\lesssim10^{-5}$ (95\% confidence), albeit with an apparent $\sim3\sigma$ preference for a nonzero value centred around $u_{\nu\chi}\sim5\times10^{-5}$~\cite{HooperLucca} (although other studies claim to rule out WDM with similar suppression~\cite{Palanque-Delabrouille:2019iyz,Gilman:2019nap}).
Given these constraints on $u_{\nu\chi}$, the interaction only occurs at appreciable rates at redshifts $z\gg1000$.
At $z\lesssim1000$, the DM particles are effectively collisionless and cold as in standard CDM, and hence amenable to standard $N$-body simulations of nonlinear structure formation.

\section{Simulating structure formation}\label{sec:galform}

Our modelling of nonlinear structure is based on a combination of semianalytic computations and $N$-body simulations.
We use the semianalytic model of structure formation \textsc{galform}, first introduced in~\cite{Cole2000}, which provides a flexible and computationally inexpensive way to produce synthetic realisations of galactic populations in a cosmological setting.
The latest version~\cite{Lacey2016} includes a large variety of astrophysical processes~\cite{Baugh2005,Bower2006,Parkinson2008,Lagos2011b} to accurately model galaxy formation and evolution.

We generate merger trees using the Monte Carlo technique described in~\cite{Parkinson2008} (which is itself based on the extended Press-Schechter (EPS) theory, and is calibrated to the results of $N$-body simulations).
In models in which the linear power spectrum $P(k)$ has a cutoff, as in our IDM scenario, a small correction is required to the EPS formalism: to obtain the variance of the density field $\sigma(M_\mathrm{h})$, $P(k)$ needs to be convolved with a sharp $k$-space filter rather than with the real-space top-hat filter used for CDM~\citep{Benson2013}.
Using our Monte Carlo technique rather than $N$-body simulations to generate merger trees has the advantage that IDM scenarios can be studied at minimum computational expense while avoiding the complication of spurious fragmentation in filaments that occurs in $N$-body simulations with a resolved cutoff in $P(k)$~(e.g.,~\cite{Wang2007,Lovell2014}).

\begin{figure}
    \centering
    \includegraphics[width=0.48\textwidth]{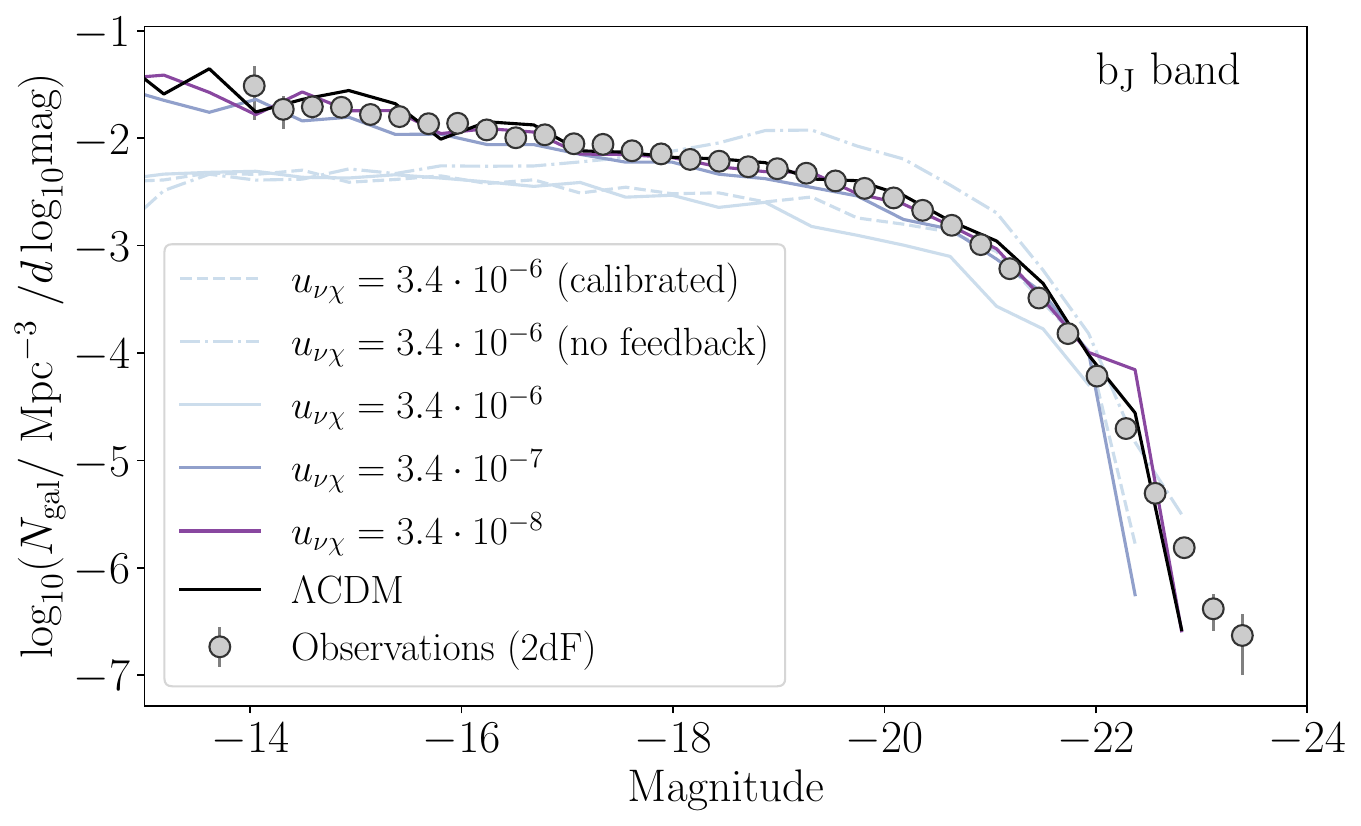}
    \caption{\label{fig:lum_func}
    Galaxy luminosity functions in the $b_J$ band for several IDM scenarios of different interaction strengths.   Observational data are from the 2dF survey~\cite{2dGRS:2001lwv}.
    It is evident that while the more weakly interaction IDM scenarios and $\Lambda$CDM are roughly consistent with observations, the $u_{\nu\chi}\sim10^{-6}$ case deviates significantly, and even in the case with feedback turned off, cannot reproduce the low-luminosity tail, despite overpredicting at high luminosities. This corresponds to a slightly weaker interaction than the preferred value of $u_{\nu\chi}=5.5^{+2.6}_{1.1}\times 10^{-6}$ from~\cite{HooperLucca}. We do not show even more strongly interacting IDM scenarios, as these deviate from observations even further.}
\end{figure}

We calibrate the astrophysical parameters in \textsc{galform} to obtain the best possible fit to observational data for $\Lambda$CDM.
In particular, we adjust parameters controlling the strength of supernova feedback to achieve the best possible match to the $z=0$ galaxy luminosity functions in the $b_J$ and $K$-bands, using observational data from~\cite{2dGRS:2001lwv,driver2012}.
We find that for IDM scenarios with $u_{\nu\chi}<10^{-6}$, the same parameters used for $\Lambda$CDM provide an equally good fit to the observational data.
For $u_{\nu\chi}\gtrsim 10^{-6}$, however, the feedback strength (as a function of halo mass) has to be reduced significantly to obtain the best possible fit.
As shown in Fig.~\ref{fig:lum_func}, this is still a worse fit than for the more weakly interaction scenarios, particularly for galaxies fainter than 19th magnitude.
We find that it is impossible for scenarios with $u_{\nu\chi}\gtrsim 10^{-5}$ to produce realistic galaxy distributions, since even with all astrophysical feedback mechanisms suppressing galaxy formation turned off, not enough galaxies are produced.
The failure to reproduce the observed galaxy luminosity functions at $z=0$ therefore already poses a stronger constraint on $u_{\nu\chi}$ than the previous tightest bounds from Lyman-$\alpha$ and rules out the preferred value of $u_{\nu\chi}=5.5^{+2.6}_{1.1}\times 10^{-6}$ found in~\cite{HooperLucca}.

\begin{figure}[t!]
    \centering
    \includegraphics[width=\linewidth]{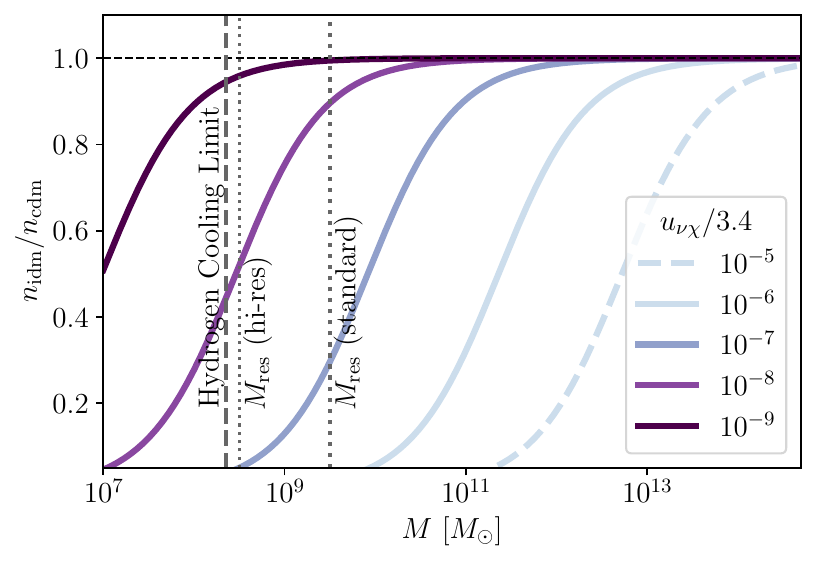}
    \caption{\label{fig:M_resolution}
    The suppression of the halo mass function, as described by Eq.~\ref{eq:suppfit}, for a range of interaction strengths.
    The dotted lines show the mass resolution of our \textsc{galform} merger trees for both the standard and high-resolution settings, while the dashed line shows the atomic hydrogen cooling limit, below which star formation is not expected~\cite{Benitez-Llambay:2020zbo}.}
\end{figure}

To ensure the validity of our \textsc{galform} results, we compute the merger trees with two different mass resolution settings, which determine the smallest tracked progenitor halos.
As shown in Fig.~\ref{fig:M_resolution}, our ``standard'' resolution is well below the minimum halo mass expected to form in IDM models with interaction strength $u_{\nu\chi}>10^{-6}$.
For the models with weaker interactions ($u_{\nu\chi}\leq10^{-7}$), the resolution is not high enough to capture the formation of the smallest halos, though it does capture a large part of the suppression in the $u_{\nu\chi}\sim10^{-7}$ case.
This motivates our choice of the ``high'' resolution halo merger trees.
Conveniently, this corresponds nearly exactly with the atomic hydrogen cooling limit, which marks the lowest mass haloes expected to form stars~\cite{Benitez-Llambay:2020zbo}.
This means that, although even our high-resolution halo merger trees don’t resolve the smallest haloes expected to form in the models with $u_{\nu\chi}<10^{-7}$, we are still able to capture the majority of the star formation expected to occur in these models.

\begin{figure}[t!]
    \centering
    \includegraphics[width=0.48\textwidth]{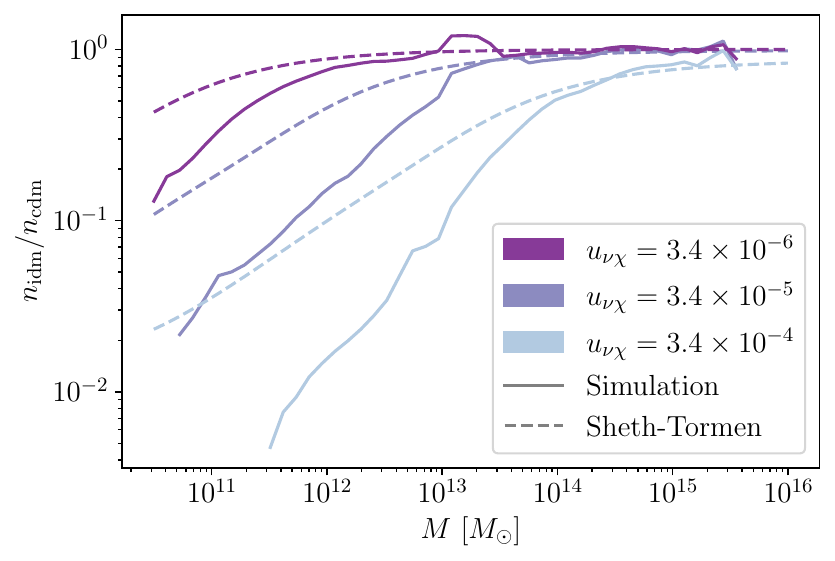}
    \caption{\label{fig:sim-vs-ST}
    IDM halo mass function $n_\mathrm{idm}$ normalised to a reference CDM mass function $n_\mathrm{cdm}$ for various interacting strengths $u_{\nu\chi}$, computed from $N$-body simulations (solid lines) and the Sheth-Tormen formalism (dashed lines).
    The choice of $u_{\nu\chi}=3.4 \times 10^{-4}$ has been motivated by the CMB+BAO constraint found in Ref.~\cite{nudm}.
    It is evident that the standard Sheth-Tormen halo mass function does not provide a good description of our simulation results.}
\end{figure}

\begin{figure*}[t!]
    \centering
    \includegraphics[width=\textwidth]{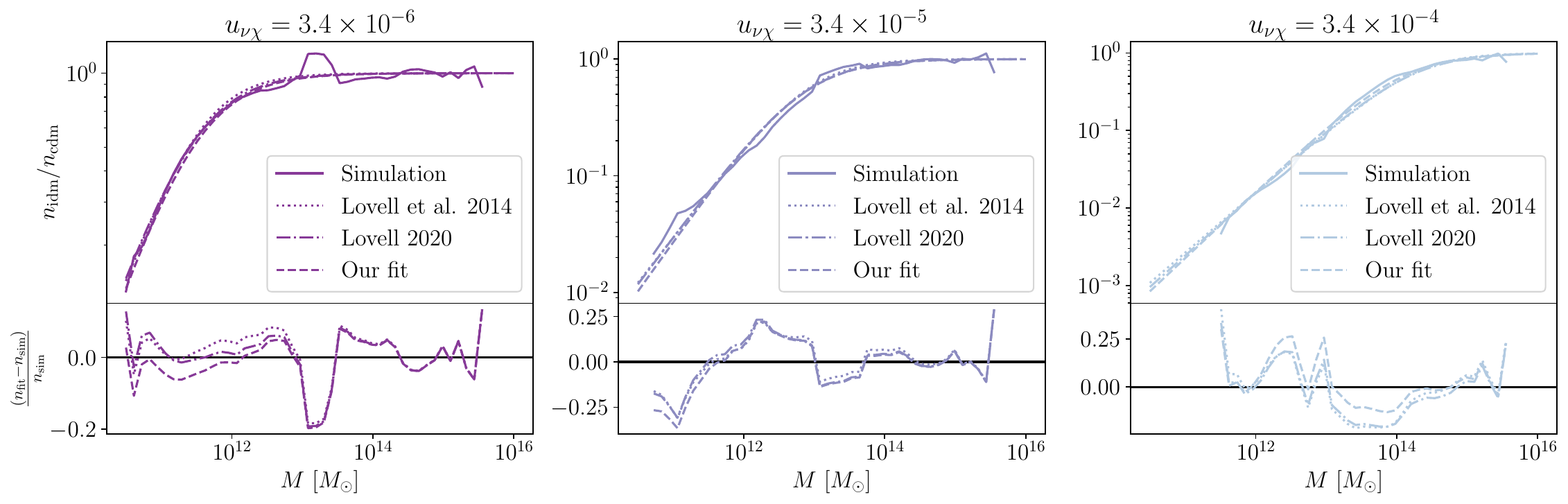}
    \caption{\label{fig:suppfit}
    IDM halo mass function $n_\mathrm{idm}$ normalised to a reference CDM mass function $n_\mathrm{cdm}$ for three interacting strengths $u_{\nu\chi}$. 
    Solid lines denote results from $N$-body simulations; dashed lines represent our fitting function~\eqref{eq:suppfit}; dotted and dot-dash lines are two fitting functions from Refs.~\cite{wdmhalos,Lovell:2020bcy}.
    The bottom panel of each plot shows the fractional residuals of the three fitting functions.}
\end{figure*}

\subsection{$N$-body simulations and the halo mass function}\label{sec:nbody}

To ensure that \textsc{galform} accurately predicts the halo mass function of the IDM scenarios, we cross-check it by performing $N$-body simulations using a version of \textsc{gadget-4}~\cite{Gadget4} modified to include the effects of massive neutrinos~\cite{supereasy} and initialised with the matter power spectra of the IDM scenarios of Fig.~\ref{fig:linear}.
We find that the \textsc{galform} and $N$-body results are compatible with each other to the $\sim\order{10\%}$ level in the mass range over which we generate merger trees using the Monte Carlo technique.

Our simulations are performed using a modified version of \textsc{gadget-4}~\cite{Gadget4}, which includes massive neutrinos via the \textsc{SuperEasy} linear response method of~\cite{supereasy}.
\textsc{gadget-4} is a massively parallel $N$-body code supporting both collisionless (gravity-only) simulations and hydrodynamical simulations for baryonic matter and star formation.
We restrict ourselves to collisionless simulations, which suffice for the purpose of cross-checking the \textsc{galform} halo mass function output. 
Initial conditions are generated at $z=49$ using the built-in \textsc{ngenic} implementation, with linear matter power spectra from our modified version of \textsc{class} as input.
For interaction strengths allowed by current cosmological data, dark matter-neutrino decoupling happens at redshifts $z \gg 1000$, well before the simulation initialisation time.
Thus, the interaction affects nonlinear structure formation only through initial conditions, and there is no need to modify the main body of \textsc{gadget-4} itself.

To probe the effects of our IDM scenario at a range of scales, we run our simulations using $N=512^3$ particles in several different box sizes, ${(50\,\text{Mpc})}^3$, ${(100\text{Mpc})}^3$, ${(500\text{Mpc})}^3$, and ${(1\text{Gpc})}^3$, for different values of the interaction strength $u_{\nu\chi}$, including a noninteracting ($u_{\nu \chi}=0$) set.
All other cosmological parameters are kept fixed to the CMB+BAO best-fit values found in~\cite{nudm}.
We also perform a set of simulations using the best-fit parameters found in~\cite{HooperLucca}, which is phenomenologically similar to our $u_{\nu\chi}=3.4\times10^{-6}$ scenario.

We use the friends-of-friends algorithm to identify haloes in our simulations.
For halo masses we use the convention $M_{200c}$, i.e., the mass contained within a sphere with mean density 200 times the critical density of the Universe. 
The haloes from each simulation are sorted into 30 evenly-spaced logarithmic mass bins to obtain the halo number density per mass bin, i.e., the halo mass function.
As is well-described in the literature (see, e.g.,~\cite{Wang2007,Lovell2014,Angulo:2013sza,Hahn:2015sia,Stucker:2021vyx}), the halo mass function of cosmologies with a small-scale cutoff exhibits unphysical discreteness effects on small scales (relative to the box size).
We therefore conservatively choose to disregard the low-mass end of each halo mass function thus constructed, cutting it off just above the scale at which discreteness effects become evident.
Then, to produce a halo mass function that spans a range of mass scales, we simply stitch together the pieces  obtained from our different box-size runs.

We have attempted to describe the IDM halo mass functions semianalytically using the Sheth-Tormen mass function, with the IDM linear matter power spectra as input and a real-space spherical tophat filter for smoothing.
However, as is evident in Fig.~\ref{fig:sim-vs-ST}, the Sheth-Tormen mass function thus constructed fails to capture the low-mass suppression relative to the CDM case seen in our $N$-body simulation results.
We therefore forego the Sheth-Tormen approach completely, opting instead to use a fitting function to describe the suppression relative to the CDM case.

We find the following fitting function to serve the purpose well:
    \begin{equation}
    \label{eq:suppfit}
        \frac{n_\text{idm}}{n_\text{cdm}}=\frac{1}{1+{\qty(\frac{M_{\beta}}{M})}^\alpha},
    \end{equation}
    where $M_\beta \equiv (4 \pi) \bar{\rho}_m/3 k_\beta^{-3}$ is the mass corresponding to the wave number $k_\beta$ at which the linear matter power spectrum of an IDM scenario is suppressed by $\beta$\% relative to the CDM case, $\bar{\rho}_m$ is the comoving mean matter density, and the parameters $\alpha$ and $\beta$ are to be determined by fitting~\eqref{eq:suppfit} to our $N$-body results.
The fit is achieved by forming and then minimising a figure-of-merit (FoM) of the form
    \begin{equation}
        \text{FoM} = \sum_M{\qty(\log_{10}{\qty(\frac{n_\text{idm}}{n_\text{cdm}})}_\text{fit}-\log_{10}{\qty(\frac{n_\text{idm}}{n_\text{cdm}})}_{N\text{-body}})}^2,
    \end{equation}
    and fitting~\eqref{eq:suppfit} to our $u_{\nu\chi}=3.4\times 10^{-4}$, $3.4 \times 10^{-5}$, and $3.4 \times 10^{-6}$ simulation results simultaneously.  

We find the best-fit values to be $\alpha=0.9$ and $\beta=10$.
The fit is displayed against simulation results in Fig.~\ref{fig:suppfit}.  We also give in Table~\ref{tab:M10} the $M_{10}$ values corresponding to several interacting strengths~$u_{\nu \chi}$.
To confirm the robustness of the fit, we test the fitting function~\eqref{eq:suppfit} against simulations at several low $u_{\nu \chi}$ values outside of the calibration range, and find the fit to be accurate to better than $\sim 10\%$.

\begin{table}[b!]
    \centering
    \begin{tabular}{l|l|c}
    $u_{\nu\chi}$     & $\:M_{10}/M_\odot$ & $\: k_{10}/h$ Mpc$^{-1}$ \\
    \hline
    $3.4\times10^{-4}\:$ & $\:1.0\times10^{14}$ & $\:0.17$ \\
    $3.4\times10^{-5}\:$ & $\:5.3\times10^{12}$ & $\:0.46$ \\
    $3.4\times10^{-6}\:$ & $\:2.3\times10^{11}$ & $\:1.3$ \\
    $3.4\times10^{-7}\:$ & $\:8.3\times10^{9}$ & $\:4.0$  \\
    $3.4\times10^{-8}\:$ & $\:2.9\times10^{8}$ & $\:12$  \\
    $3.4\times10^{-9}\:$ & $\:9.6\times10^{6}$ & $\:38$ \\
    \end{tabular}
    \caption{IDM interacting strengths and the corresponding suppression mass scale $M_{\beta}$ that appears in the fitting function~\eqref{eq:suppfit} for the halo mass function.  We find $\beta=10$ to be the best-fit value against simulation results.
    $k_{10}$ is the corresponding $k$-scale.\label{tab:M10}}
\end{table}

Finally, we remark that extended Press-Schechter formalism used in \textsc{galform} does not exactly use our fitting function, but produces a halo mass function that matches it to the same or better precision as the fitting function matches the simulations.

\section{Gravitational-wave event rates}\label{sec:gravwaves}

Our key observable is the redshift-dependent BBH merger rate density $\mathcal{R}_\mathrm{BBH}(z)$.
In principle we could also use our simulations to study the merger rates of binary neutron stars and black hole-neutron star binaries; however, these are only detectable at lower redshifts due to their smaller masses and correspondingly weaker gravitational-wave signals.
They are thus less sensitive as a probe of the high-redshift suppression effects we are interested in.

Stellar-mass BBH mergers are thought to form primarily through isolated binary evolution, in which massive binary stars end their lives as black holes and radiate orbital energy through gravitational-wave emission until they eventually merge, generating an observable gravitational-wave event~\cite{Mandel:2018hfr}.
The BBH merger rate is thus essentially a delayed tracer of star formation, whose normalisation depends on the efficiency with which massive binary stars are converted into BBHs.
This efficiency is mostly determined by the stellar metallicity (i.e., the fraction of the stellar mass that is in elements heavier than helium).

We model the merger rate using the binary population synthesis code \textsc{compas}~\cite{Compas,Stevenson:2017tfq,Vigna-Gomez:2018dza}, which generates a synthetic BBH population by sampling individual stellar binaries from initial mass and metallicity distributions and evolving them over cosmic time.
\textsc{compas} tracks the stellar and orbital evolution of each binary, accounting for a range of physical processes such as stellar winds, mass transfer, BH formation at the end of each star's lifetime, and subsequent shrinking of the orbit via gravitational-wave radiation reaction.
We use a \textsc{compas} dataset of 20 million evolved binaries (resulting in $\approx0.7$ million BBHs) presented in~\cite{Riley:2020btf}, which is publicly available at~\cite{riley_jeff_2021_5595426}.
This gives us the BBH formation efficiency as a function of initial mass and metallicity, as well as the delay time between star formation and BBH merger.
By combining this with a model for the star formation rate density and metallicity distribution as functions of redshift, we can use the \textsc{compas} ``cosmic integration'' module~\cite{Neijssel:2019irh} to average over the synthetic population and obtain the cosmic BBH merger rate.
These inputs are typically phenomenological models chosen to fit low-redshift observational data.
In our case, we instead use the star formation rates and metallicity distributions generated by \textsc{galform}, allowing us to model how the BBH rate changes as we vary the underlying dark matter model.

\begin{figure}[t!]
    \centering
    \includegraphics[width=0.48\textwidth]{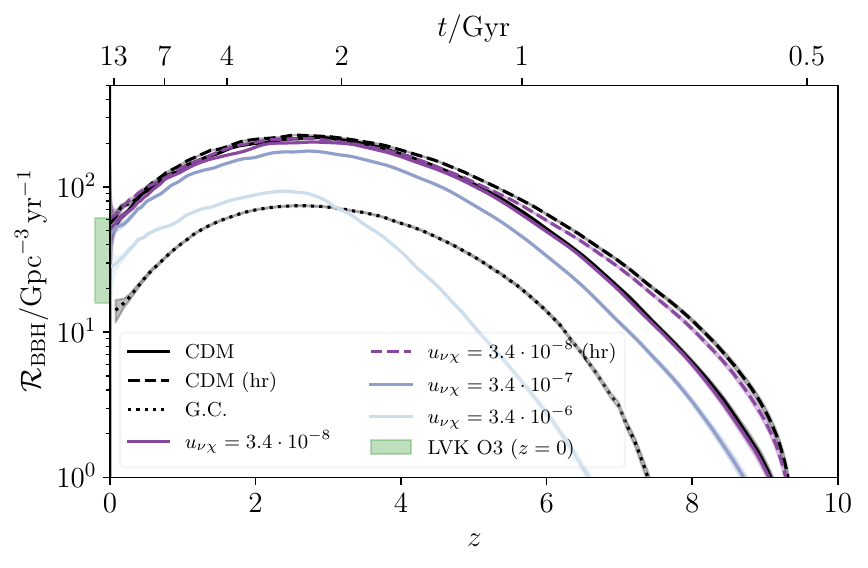}
    \caption{\label{fig:merger-rates}
    The binary black hole merger rate density over cosmic time, as predicted by our pipeline for $\Lambda$CDM and a range of IDM scenarios.
    The local $z=0$ rate ($90\%$ C.L.) inferred from LIGO/Virgo observations~\cite{KAGRA:2021duu} is shown as a green band.
    Curves marked ``hr'' are high resolution, meaning that the \textsc{galform} merger trees are tracked to smaller progenitor masses.
    The curve marked ``G.C.'' is the contribution from globular clusters in the noninteracting scenario.}
\end{figure}

Our key results for the BBH merger rate are shown in Fig.~\ref{fig:merger-rates}.
In addition to the baseline $\Lambda$CDM case, we consider a range of IDM models with $u_{\nu\chi}=3.4\times\{10^{-6},10^{-7},10^{-8}\}$ (recall that models with stronger DM-$\nu$ interactions than this are unable to reproduce the galaxy luminosity function data shown in Fig.~\ref{fig:lum_func}).
The factor of $3.4$ comes from the CMB+BAO limits in~\cite{nudm}; we have chosen interaction strength values that are whole factors of 10 smaller.
The rates that we obtain depend on the mass resolution of our \textsc{galform} merger trees, with finer resolution allowing us to access additional star formation taking place in smaller haloes, resulting in a greater total number of BBHs.
Most of the results shown in Fig.~\ref{fig:merger-rates} use a benchmark mass resolution of $M_\mathrm{h}\ge10^{9.5}\,M_\odot$, but we additionally show results for CDM and for $u_{\nu\chi}=3.4\times10^{-8}$ at a finer resolution of $M_\mathrm{h}\ge10^{8.5}\,M_\odot$.
At this mass resolution, we are able to resolve the majority of the star formation in the Universe (under the assumption that it takes place through the cooling of atomic hydrogen), except for that from the ultrafaint dwarf population.
We do not anticipate these galaxies to contribute significantly to the overall BBH merger rate.

One limitation of our approach is that \textsc{compas} only accounts for BBHs formed through isolated binary evolution, neglecting other formation channels such as dynamical capture in dense stellar environments~\cite{LIGOScientific:2010nhs,Mandel:2018hfr} (globular clusters~\cite{OLeary:2007iqe}, nuclear star clusters~\cite{Miller:2008yw}, etc.) or in AGN disks~\cite{Mckernan:2017ssq}, as well as the possible presence of primordial BBHs~\cite{Bird:2016dcv}.
However, these additional merger channels can be safely neglected in a first analysis, as they are expected to be subdominant compared to the contribution from isolated binary evolution.
For example, Fig.~\ref{fig:merger-rates} shows the contribution from dynamical BBH assembly in globular clusters, as calculated in~\cite{Rodriguez:2018rmd} by combining cluster $N$-body simulations~\cite{Rodriguez:2017pec} with a semianalytical cosmological model of globular cluster abundances~\cite{2019MNRAS.482.4528E}.
We see that this contribution is subdominant at all redshifts, justifying our focus here on isolated binary evolution.
In principle, however, one could incorporate subdominant channels such as these into our analysis with additional modelling, e.g., so that we capture the suppression of globular cluster abundances due to IDM.

\begin{figure}[b!]
    \centering
    \includegraphics[width=\linewidth]{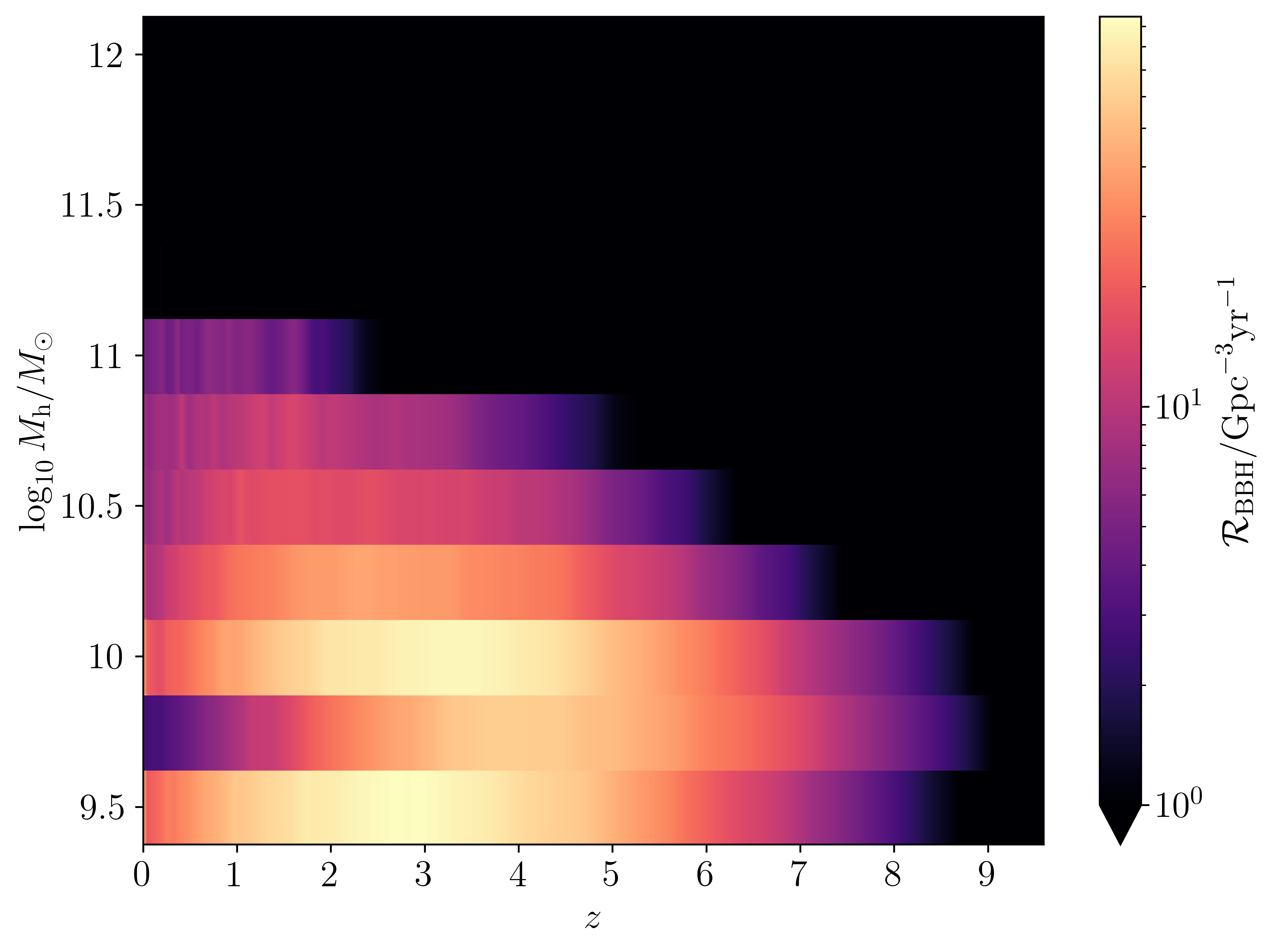}
    \includegraphics[width=\linewidth]{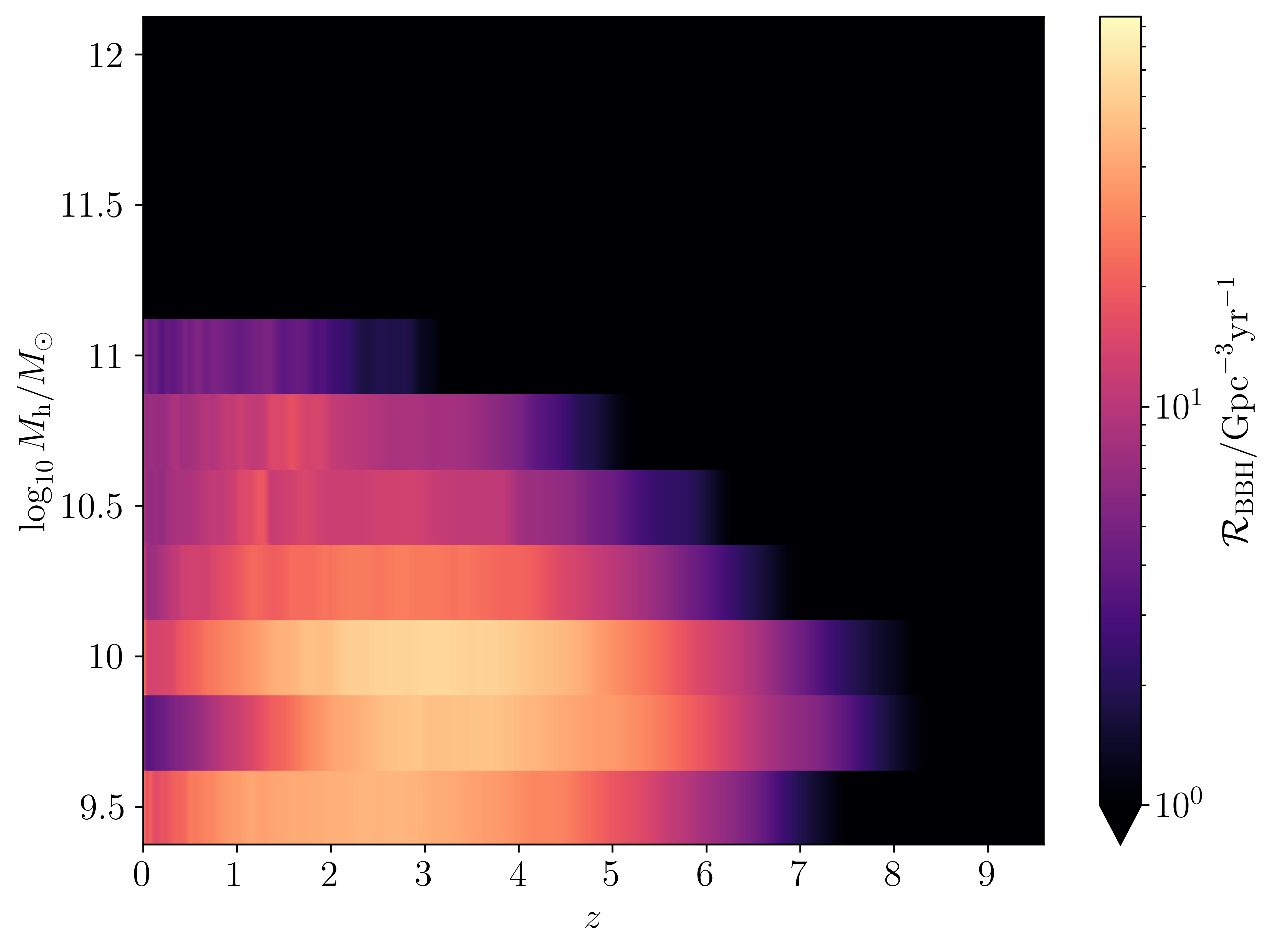}
    \caption{\label{fig:mhalo-bins}
    BBH merger rate as a function of redshift and host halo mass, in both the (standard-resolution) CDM case and the $u_{\nu\chi}\sim10^{-7}$ case.}
\end{figure}

\subsection{BBH merger rate as a function of host halo mass}

The advantage of our proposed method for constraining dark matter models is that it allows us to target the high-redshift, low-mass haloes that are most strongly suppressed in these models.
This is highlighted in Fig.~\ref{fig:mhalo-bins}, where we break the total redshift-dependent BBH merger rate down into bins in halo mass for CDM and for an IDM model with $u_{\nu\chi}\sim10^{-7}$.
We see that the strongest contribution to the observed merger rate, across all redshifts, comes from light haloes, $M_\mathrm{h}\lesssim10^{10}\,M_\odot$.
These are precisely the haloes that are suppressed for DM-$\nu$ interactions on the order of $u_{\nu\chi}\sim10^{-7}$ (cf. Fig.~\ref{fig:M_resolution}), which explains why the BBH merger rate is a useful probe of this suppression.

In particular, we see in Fig.~\ref{fig:mhalo-bins} that the suppression is most apparent at high redshifts, $z\gtrsim7$.
This again highlights the advantages of our method, as galaxies in these light haloes are challenging to observe directly at such high redshift, even with JWST.
For instance, at $z\gtrsim7$ the observational limit for JWST assuming a Hubble Ultra Deep Field-like survey is around $M_\star\sim10^{7.5}\,M_\odot$, or $M_\mathrm{h}\sim10^{9-10}\,M_\odot$~\cite{Yung:2019,Khimey:2021} depending on the stellar-to-halo mass relation assumed.
Even under the assumption of this optimistic scenario, the BBH merger rate technique allows us, in principle, to probe halo masses one to two orders of magnitude lower.

\subsection{Future gravitational-wave detection forecasts}

\begin{figure}[t!]
    \centering
    \includegraphics[width=0.48\textwidth]{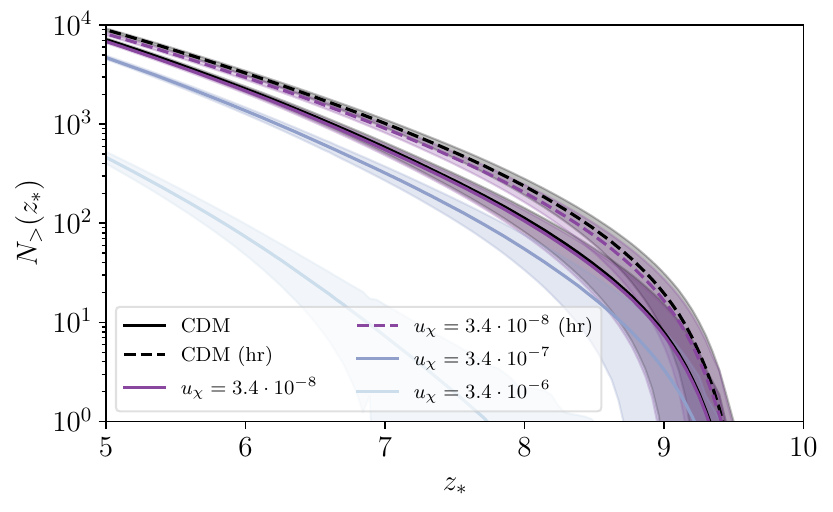}
    \caption{\label{fig:ncut}
    Expected number of BBH mergers observed with estimated redshift larger than $z_*$, using one year of observations with a third-generation interferometer network (Einstein Telescope plus two Cosmic Explorers), taking into account the forecast redshift uncertainty and signal-to-noise for the expected BBH population.
    The shaded regions show the statistical uncertainty ($99\%$ C.L.) due to Poisson fluctuations in the number of BBHs, and `hr' means high resolution, signifying that the \textsc{galform} merger trees are tracked to smaller progenitor masses.}
\end{figure}

In order to convince ourselves that differences in the high-$z$ BBH merger rate will be detectable with future gravitational-wave observatories, we compute forecasts for the expected number of BBHs, $N_>(z_*)$, detected above a given threshold redshift $z_*$, as well as the expected uncertainty in this quantity due to Poisson fluctuations in the number of BBHs at each redshift, as well as redshift measurement errors.
(This is based on the ``cut-and-count'' method discussed in~\cite{Martinelli:2022elq}.)
We assume one year of observations with a third-generation interferometer network consisting of Einstein Telescope~\cite{Maggiore:2019uih} and two Cosmic Explorers~\cite{Evans:2021gyd}, accounting for the detection efficiency and redshift uncertainty associated with this network.
As shown in Fig.~\ref{fig:ncut}, we can clearly distinguish between the different $N_>(z_*)$ predictions, allowing us to confirm or rule out a small-scale suppression of the scale caused by DM-neutrino interactions down to the level of $u_{\nu\chi}\sim10^{-7}$.

Figure~\ref{fig:ncut} clearly demonstrates that the main limitation to the cosmological information we can extract from future gravitational-wave detections is not statistical uncertainty, but systematics associated with modelling choices.
In particular, since the observational study of BBHs is still at an early stage, there are numerous astrophysical uncertainties associated with the stellar physics that governs isolated binary evolution.
We return to this issue later in this section, where we investigate the impact of these uncertainties and show that they are not degenerate with the DM suppression effect.

\begin{figure*}[t!]
    \centering
    \includegraphics[width=0.497\textwidth]{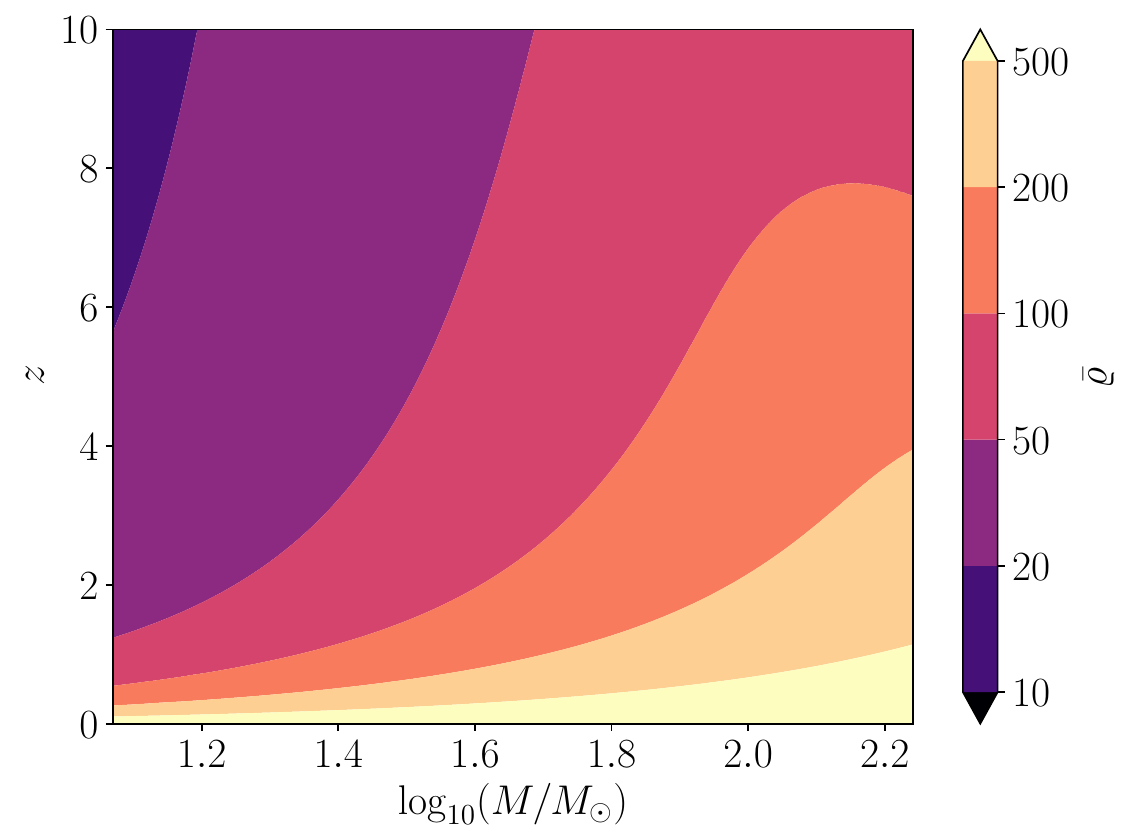}
    \includegraphics[width=0.497\textwidth]{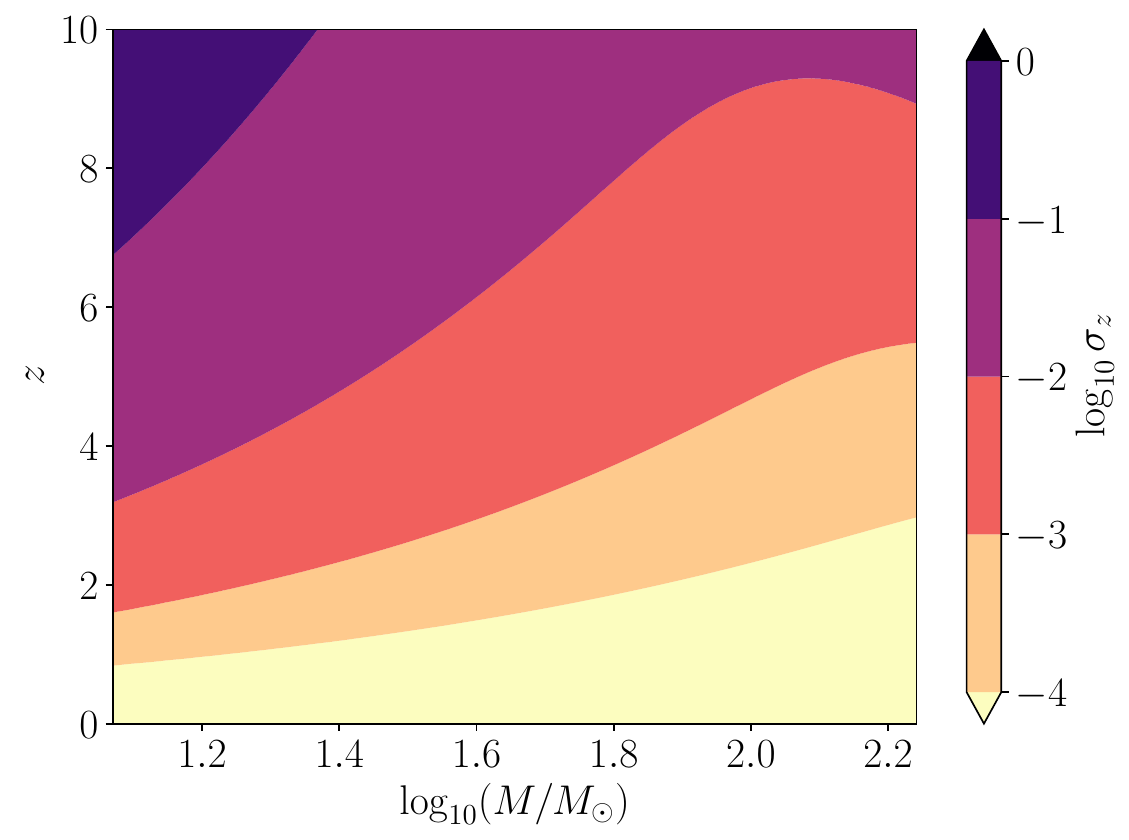}
    \caption{\label{fig:forecasts}
    Left panel: mean signal-to-noise ratio $\bar{\varrho}=\sqrt{(h|h)}$ of equal-mass nonspinning BBHs detected with our third-generation interferometer network, as a function of redshift $z$ and total source-frame mass $M$.
    Right panel: forecast redshift uncertainty $\sigma_z=\sqrt{{(F^{-1})}_{zz}}$ for the same parameter space.
    The mass range in both panels is truncated at the boundaries of our broken-power-law mass distribution ($12\,M_\odot<M<174\,M_\odot$).}
\end{figure*}

Our key observable quantity is constructed as follows: we choose some redshift threshold $z_*$, and define $N_>(z_*)$ as the number of detected BBHs whose best-fit inferred redshift falls above this threshold, $\hat{z}>z_*$.
Our results show that different dark matter scenarios will predict different values of $N_>$.
We can predict whether these models are distinguishable in a given observational scenario by forecasting the corresponding uncertainty on $N_>$.
It is straightforward to write down the mean value of this quantity,
    \begin{align}
    \begin{split}
    \label{eq:count-mean}
        \bar{N}_>(z_*)=T_\mathrm{obs}&\int_0^\infty\dd{z}\dv{V}{z}\frac{\bar{\mathcal{R}}(z)}{1+z}\\
        &\times\int\dd{\vb*\theta}p_\mathrm{pop}(\vb*\theta|z)P_{\det}(z,\vb*\theta)P_>(z,\vb*\theta|z_*).
    \end{split}
    \end{align}
Here, $T_\mathrm{obs}$ is the total observing time, $V$ is the comoving volume as a function of redshift, $\bar{\mathcal{R}}$ is the mean BBH merger rate density as a function of redshift, $\vb*\theta$ is a vector of BBH parameters (masses, spins, etc.) with population distribution $p_\mathrm{pop}(\vb*\theta|z)$ (which we allow to evolve over redshift), $P_{\det}(z,\vb*\theta)$ is the probability of detecting a BBH with parameters $\vb*\theta$ at redshift $z$, and $P_>(z,\vb*\theta|z_*)$ is the probability of inferring a best-fit redshift greater than $z_*$ for the same BBH, the width of which encapsulates redshift measurement uncertainties for a given GW detector network.
We describe how to calculate each of these probabilities below.

In order to calculate the variance of $N_>$, we assume that the number, $n$, of BBH signals emitted from a comoving volume $V$ in a source-frame time $T=T_\mathrm{obs}/(1+z)$ is a Poisson random variable with mean $\bar{n}=VT\bar{\mathcal{R}}$.
We further assume that Poissonian fluctuations in the BBH rate at different redshifts are uncorrelated with each other.
This allows us to write the covariance matrix between these BBH number counts as
    \begin{equation}
        \mathrm{Cov}[n(z),n(z')]=\delta(z-z')\mathrm{Var}[n(z)]=\delta(z-z')\bar{n}(z).
    \end{equation}
We thus find that
    \begin{align}
    \begin{split}
    \label{eq:count-var}
        \mathrm{Var}[N_>(z_*)]&=T_\mathrm{obs}\int_0^\infty\dd{z}\dv{V}{z}\frac{\bar{\mathcal{R}}(z)}{1+z}\\
        &\times{\qty[\int\dd{\vb*\theta}p_\mathrm{pop}(\vb*\theta|z)P_{\det}(z,\vb*\theta)P_>(z,\vb*\theta|z_*)]}^2,
    \end{split}
    \end{align}
    which shows that $N_>$ is \emph{not} itself Poissonian, since its mean and variance are unequal so long as the $\vb*\theta$ integral is not equal to unity.

We assume the BBH signals are detected using a matched-filter search, in which the detection statistic is a noise-weighted inner product of the data $d$ with the signal template $h$,
    \begin{equation}
        (d,h)\equiv4\Re\int_0^\infty\dd{f}\frac{\tilde{d}(f)\tilde{h}^*(f)}{S(f)},
    \end{equation}
    with $S(f)$ the noise power spectral density of the detector.
In Gaussian noise, the signal-to-noise ratio (SNR) $\varrho$ of the search is then approximately a Gaussian random variable with unit variance and mean $\bar{\varrho}=\sqrt{(h,h)}$; so for example, having $\bar{\varrho}=5$ would constitute a $5\sigma$ detection in this idealised scenario.
In reality, the noise is not Gaussian, and we require a larger SNR in order to confidently discriminate a BBH from various non-Gaussian noise transients.
A commonly adopted detection threshold is $\bar{\varrho}\ge8$.
We can therefore write the detection probability as
    \begin{align}
    \begin{split}
        P_{\det}(z,\vb*\theta)&=\int_8^\infty\frac{\dd\varrho}{\sqrt{2\uppi}}\exp[-\frac{1}{2}{\qty(\varrho-\bar{\varrho}(z,\vb*\theta))}^2]\\
        &=\frac{1}{2}\qty[1+\erf\qty(\frac{\bar{\varrho}(z,\vb*\theta)-8}{\sqrt{2}})],
    \end{split}
    \end{align}
    which interpolates between (almost) zero for faint signals and (almost) one for strong signals, with a characteristic width set by the randomness of the observed SNR.

We approximate the redshift inference using a Fisher forecast.
The Fisher matrix for parameters $\theta_i$ (including $z$) of the signal model $h(\theta_i)$ is given by
    \begin{equation}
        F_{ij}\equiv\qty(\pdv{h}{\theta_i},\pdv{h}{\theta_j}).
    \end{equation}
In the strong-signal limit, the inferred redshift $\hat{z}$ is a Gaussian random variable with mean equal to the true redshift $z$, and standard deviation given by the inverse Fisher matrix,
    \begin{equation}
        \sigma_z(z,\vb*\theta)=\sqrt{{(F^{-1})}_{zz}}.
    \end{equation}
We therefore have
    \begin{equation}
        P_>(z,\vb*\theta|z_*)=\frac{1}{2}\qty[1+\erf\qty(\frac{z-z_*}{\sqrt{2}\sigma_z})],
    \end{equation}
    which interpolates between (almost) zero for nearby BBHs and (almost) one for distant BBHs, with a width set by~$\sigma_z$.

In order to evaluate the expected SNR $\bar{\varrho}$ and the redshift uncertainty $\sigma_z$, we need a signal model $h(\vb*\theta,z)$.
For this, we use the phenomenological hybrid waveform model of~\cite{Ajith:2007kx}.
For simplicity, we assume that the BBHs are all equal-mass and nonspinning (as is approximately true for most of the BBHs detected by LIGO/Virgo thus far~\cite{KAGRA:2021duu}), and we average out all of the extrinsic parameters (sky location, polarisation angle, inclination, etc.).
This reduces the number of relevant parameters to just two: the redshift $z$ and the total mass $M$.
In terms of these parameters, the frequency-domain waveform model can be written as
    \begin{align}
    \begin{split}
        \tilde{h}(f)&=\rme^{\rmi\Psi(f)}{\qty(\frac{\uppi}{30})}^{1/2}\frac{{(GM)}^2}{2r}{(\uppi GMf_\mathrm{merge})}^{-7/6}\\
        &\times
        \begin{cases}
            {(f/f_\mathrm{merge})}^{-7/6} & f<f_\mathrm{merge} \\
            {(f/f_\mathrm{merge})}^{-2/3} & f_\mathrm{merge}\le f<f_\mathrm{ring} \\
            \frac{{(f_\mathrm{ring}/f_\mathrm{merge})}^{-2/3}f_\mathrm{width}^2}{{(f-f_\mathrm{ring})}^2+f_\mathrm{width}^2} & f_\mathrm{ring}\le f<f_\mathrm{cut}
        \end{cases},\\
        \Psi(f)&=\varphi_0+\sum_{k=0}^7\psi_k{(\uppi GMf)}^{(k-5)/3},
    \end{split}
    \end{align}
    where $f_\mathrm{merge}$, $f_\mathrm{ring}$, $f_\mathrm{cut}$, $f_\mathrm{width}$, $\varphi_0$, and $\{\psi_k\}$ are all constants given in~\cite{Ajith:2007kx}.
This expression is appropriate for $z\to0$; in order to generalise to arbitrary redshift we simply replace $r\to d_L(z)$ (with $d_L$ the luminosity distance), $M\to(1+z)M$, and $f\to f/(1+z)$ for all of the frequencies.

We use this waveform to compute the expected SNR $\bar{\varrho}$ and the redshift uncertainty $\sigma_z$ (the latter of which involves computing and inverting the $2\times2$ Fisher matrix for the mass and redshift), as shown in Fig.~\ref{fig:forecasts}.
The final missing ingredient to calculate $N_>$ and its variance is the population distribution of the BBH masses, for which we use the best-fit broken-power-law model from~\cite{LIGOScientific:2020kqk}, which transitions from a shallow $m^{-1.58}$ power law at low masses ($5.9\,M_\odot<m<41\,M_\odot$) to a steeper one, $m^{-5.59}$, at large masses ($41\,M_\odot<m<87\,M_\odot$).
Here $m$ is the source-frame mass of each individual black hole, so $M=2(1+z)m$.

\begin{figure*}[t!]
    \centering
    \includegraphics[width=0.497\textwidth]{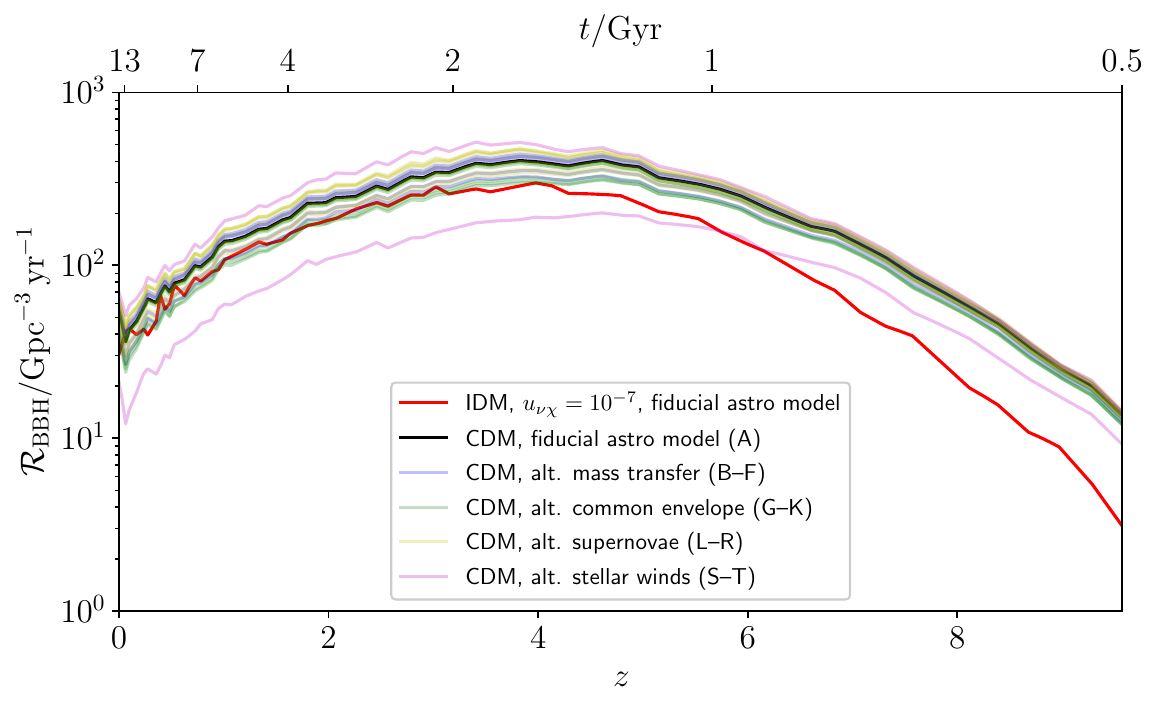}
    \includegraphics[width=0.497\textwidth]{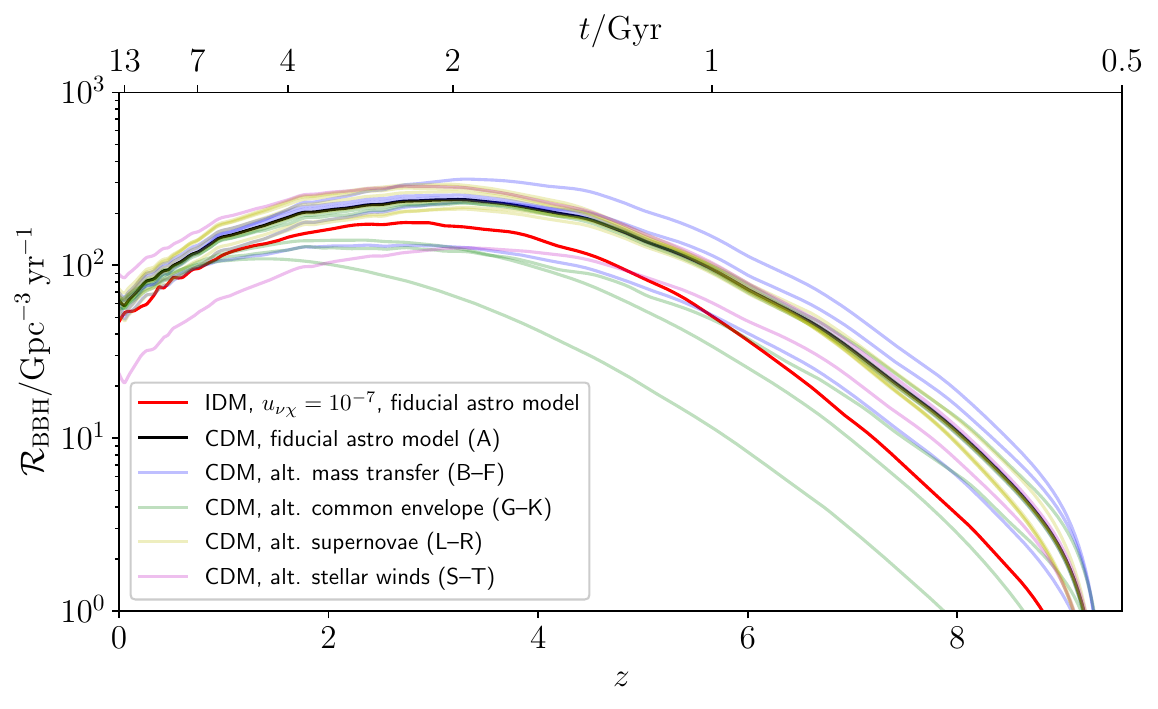}
    \caption{\label{fig:rates-comparison}
    Formation rate (left panel) and merger rate (right panel) of BBHs in 21 different scenarios: a fiducial CDM model, an IDM model with $u_{\nu\chi}\sim10^{-7}$ and fiducial astrophysical settings, and 19 alternative CDM models in which various \textsc{compas} parameters are varied.}
\end{figure*}

\subsection{Varying astrophysical model parameters}

While our results show that the high-$z$ BBH merger rate is sensitive to DM microphysics, it undoubtedly also depends on the astrophysical modelling choices we have made in our pipeline.
In order to investigate the extent to which variations in these choices are degenerate with the DM suppression effect, we use the \textsc{compas} simulation data from~\cite{Broekgaarden:2021efa} (publicly available at~\cite{floor_broekgaarden_2021_5651073}), consisting of 19 alternative models in which the binary stellar evolution modelling is systematically modified, isolating the effects of mass transfer, common envelopes, supernovae, and stellar winds.
We focus on modifications to \textsc{compas} modelling choices rather than those in \textsc{galform}, as the latter are currently much more tightly constrained by fitting to observational data.
In fact, the predictive power of the \textsc{galform} model is enhanced by the fact that the same set of parameters that show consistency with low redshift observational data also predict reasonably accurate galaxy abundances at high redshift (relevant to the epochs of interest in this paper) across multiple wavelengths~\cite{Hou:2016,Cowley:2018,Cowley:2019}.

In Fig.~\ref{fig:rates-comparison} we show the BBH formation rate and merger rate in 21 different scenarios: our fiducial CDM model (``A''), 19 alternative CDM models (``B''--``T'') with alternative modelling of binary stellar evolution, and a fiducial IDM model with $u_{\nu\chi}=3.4\times10^{-7}$.
We see that the IDM model is clearly distinguished from the other 20 models in terms of its BBH \emph{formation} rate, with a characteristic high-$z$ suppression relative to these other models.
However, the distinction between IDM and the other alternative models is blurred when considering the BBH \emph{merger} rate, which we can understand as being due to the way these models modify the distribution of delay times between formation and merger.

Nonetheless, a more detailed analysis reveals that IDM is still distinguishable from the alternative models, even if we only have access to the BBH merger rate.
To demonstrate this, we define a parameterised merger rate model,
    \begin{equation}
    \label{eq:parameterised-model}
        \mathcal{R}(z,\{\theta_i\})=\mathcal{R}_\mathrm{A}(z)+\sum_i\theta_i[\mathcal{R}_i(z)-\mathcal{R}_\mathrm{A}(z)],
    \end{equation}
    where the index $i$ labels models IDM, B, C, \ldots, T.
The case where $\theta_i=0$ for all $i$ corresponds to our fiducial model A, while having $\mathcal{R}$ equal to a particular alternative model corresponds to $\theta_i=1$ for that case and $=0$ otherwise.
Between these special points in parameter space,~\eqref{eq:parameterised-model} linearly interpolates between the 21 different models.
We can therefore investigate the degeneracy between the different models by carrying out a Fisher analysis on the $\theta_i$ parameters.
We stress that this is not intended to represent a realistic model of the BBH merger rate over the full space of astrophysical model parameters (this would require us to carry out simulations for a large number of points sampled from the 20+ dimensional parameter space, which is prohibitively expensive), but rather to provide an approximate tool for assessing the degeneracy of IDM suppression with astrophysical modelling choices.

To perform our Fisher analysis for the parameters $\{\theta_i\}$, we imagine sorting a large number of observed BBHs into redshift bins, and counting the number $N(z)$ in each bin.\footnote{%
    These number counts are related to the ``cut-and-count'' quantity $N_>$ from the previous section by $N(z)=N_>(z_{\min})-N_>(z_{\max})$, where $z_{\max}$, $z_{\min}$ are the upper and lower edges of bin $z$.}
We model these counts with a Gaussian likelihood,
    \begin{equation}
        -2\mathcal{L}(N|\{\theta_i\})=\sum_{z\in\mathrm{bins}}\frac{1}{\sigma_z^2}{\qty[N(z)-\bar{N}(z,\{\theta_i\})]}^2+\mathrm{const.},
    \end{equation}
    where the mean count in each bin, $\bar{N}(z)$, is calculated from~\eqref{eq:count-mean} using the parameterised merger rate~\eqref{eq:parameterised-model}, and the variance $\sigma_z^2$ is calculated from~\eqref{eq:count-var} using the fiducial rate $\mathcal{R}_\mathrm{A}$.
(Recall that this variance includes redshift measurement errors.)
The Fisher matrix is then
    \begin{equation}
    \label{eq:fisher-degeneracy}
        F_{ij}=\sum_{z\in\mathrm{bins}}\frac{1}{\sigma_z^2}\qty[\bar{N}_i(z)-\bar{N}_\mathrm{A}(z)]\qty[\bar{N}_j(z)-\bar{N}_\mathrm{A}(z)].
    \end{equation}

\begin{figure}[t!]
    \centering
    \includegraphics[width=0.497\textwidth]{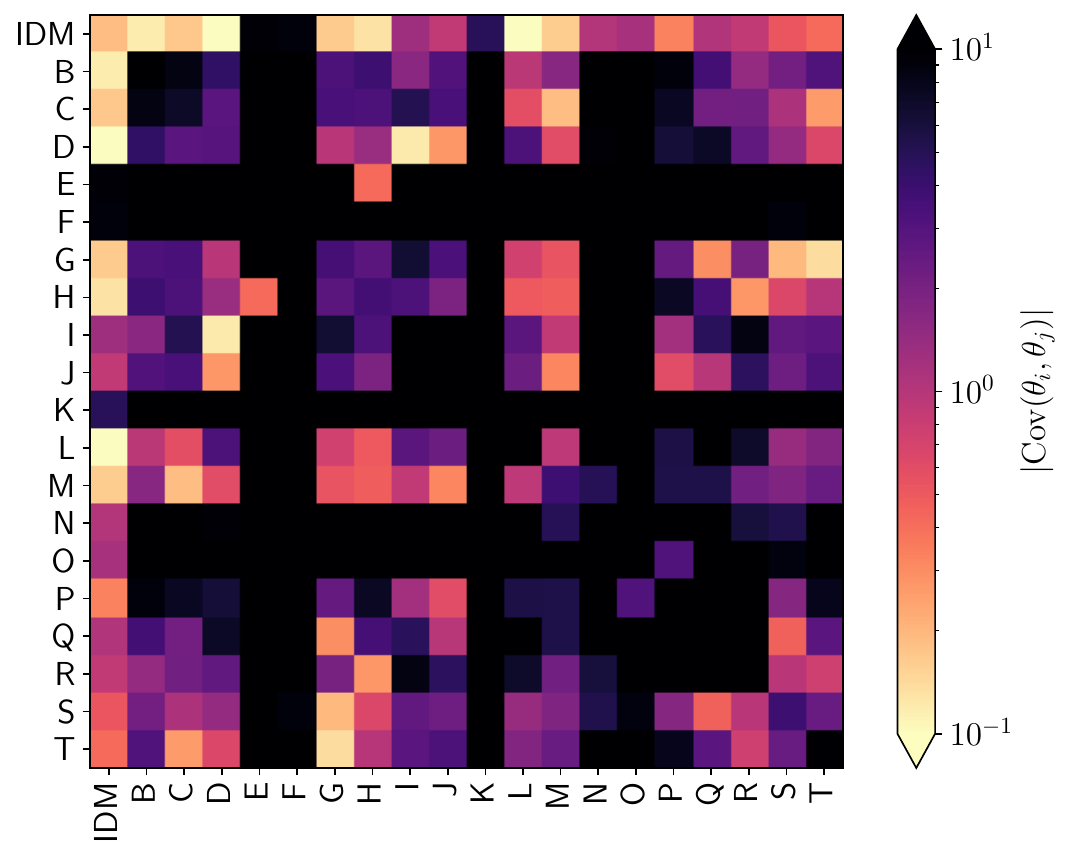}
    \caption{\label{fig:degeneracy}
    Covariance of the $\theta_i$ parameters defined in~\eqref{eq:parameterised-model} based on the Gaussian Fisher matrix~\eqref{eq:fisher-degeneracy}, assuming one year of observations with a third-generation GW interferometer network.}
\end{figure}

In Fig.~\ref{fig:degeneracy} we show the covariance matrix, $\mathrm{Cov}(\theta_i,\theta_j)={(F^{-1})}_{ij}$, computed from~\eqref{eq:fisher-degeneracy} using 38 redshift bins of width $\upDelta z=0.25$ over the range $z\in[0,9.25]$, assuming one year of observations with a third-generation GW interferometer network as described in the main text.
We are particularly interested in the on-diagonal elements of this matrix, which indicate the measurement variance for each of the $\theta_i$; roughly speaking, a variance significantly smaller than unity for a given model indicates that the BBH merger rate in that model can be distinguished from all other models with one year of data.
We find that $\mathrm{Var}(\theta_\mathrm{IDM})\approx0.18$, so that the DM suppression effect is \emph{not} degenerate with the other models considered here, at least for an interaction strength of $u_{\nu\chi}\sim10^{-7}$; in fact, IDM is the only model in which this is the case, indicating that the alternative models B--T are all degenerate with each other.
This variance corresponds to a $\sim2.3\sigma$ detection of the IDM suppression after one year of observations.
Since the Fisher information grows linearly with observation time (see~\eqref{eq:count-mean} and~\eqref{eq:count-var}), this implies a $5\sigma$ detection with $\sim4.6\,\mathrm{yr}$ of data.

\section{Summary and conclusion}\label{sec:summary}

We have investigated the suppression of the merger rate of binary black holes in scenarios where dark matter scatters off of neutrinos, taking this as an example of a broader family of dark matter models in which structure is suppressed on small scales.
Our predictions are based on a simulation pipeline (cf. Fig.~\ref{fig:flowchart}) that captures all of the crucial physical ingredients of the problem, from the initial (linear) matter power spectra, through the formation and evolution of galaxies, to a population of merging BBHs.
Our key results are shown in Figs.~\ref{fig:merger-rates} and~\ref{fig:ncut}, which respectively show the suppression of the BBH merger rate and demonstrate that this suppression will be detectable with next-generation gravitational-wave observatories.

All of the dark matter scenarios we consider are consistent with the local ($z=0$) merger rate inferred by LIGO/Virgo, $\mathcal{R}_\mathrm{BBH}(z=0)=[16\text{--}61]\,\mathrm{Gpc}^{-3}\,\mathrm{yr}^{-1}$~\cite{KAGRA:2021duu}, with only a small suppression ($\lesssim20\%$) of this local rate in the $u_{\nu\chi}\le3.4\times10^{-7}$ IDM scenarios compared to $\Lambda$CDM.
At higher redshifts, however, we see that their differences become increasingly significant.
Physically this reflects that most star formation takes place in smaller haloes at these early epochs, and these smaller haloes are where the effects of DM interactions are most pronounced.
Even for interactions as weak as $u_{\nu\chi}=3.4\times10^{-8}$ there is a small but visible difference from CDM in the predictions from our high-resolution simulations.
Decreasing the interaction strength further down to $u_{\nu\chi}=3.4\times10^{-9}$, however, makes the suppression scale so small that there are very few star-forming haloes that are missing compared to $\Lambda$CDM, and consequently the BBH merger rate for this scenario is essentially indistinguishable from that in $\Lambda$CDM.

To confirm that the different dark matter models can indeed be distinguished in this way, we have carried out a Fisher forecast for the redshift-dependent BBH number count $N_>(z_*)$, as shown in Fig~\ref{fig:ncut}.
We find that future observations with Einstein Telescope and Cosmic Explorer will allow us to confirm or rule out suppression caused by DM-neutrino interactions down to the level of $u_{\nu\chi}\sim10^{-7}$.

The main challenge for this method is the significant astrophysical uncertainties in the modelling.
We stress however that there is good reason to be optimistic regarding these systematic uncertainties: any changes to the binary evolution model will have an impact at \emph{all} redshifts and should therefore have minimal degeneracy with the specific high-redshift suppression we are targeting.
We envisage that the very large numbers of BBHs detected by third-generation interferometers at low redshifts will be able to pin down the astrophysical modelling, allowing us to use the high-redshift tail as a sensitive probe of structure formation and dark matter.
We demonstrate this explicitly by investigating 19 alternative \textsc{compas} models~\cite{Broekgaarden:2021efa}; we find that the suppression effect from IDM (at the level of $u_{\nu\chi}\sim10^{-7}$) can be confidently distinguished from each of these alternatives with next-generation GW observations.
Similarly, we anticipate that uncertainties associated with the \textsc{galform} modelling governing star formation and chemical enrichment will be reduced by comparing against future low-redshift observational data for faint galaxies, as well as high-redshift observations with, e.g., the James Webb Space Telescope~\cite{Gardner:2006ky}, leaving little degeneracy with DM microphysics.

To conclude, we have shown that the binary black hole merger rate has the potential to become an important probe of deviations from the $\Lambda$CDM model, particularly the suppression of structure from dark matter interactions.
As a by-product of our analysis, we have shown that dark matter-neutrino interactions with $u_{\nu\chi}\gtrsim10^{-6}$ are already strongly in tension with the observed abundance of faint galaxies.
With the next generation of gravitational-wave observatories probing the gravitational-wave event rate out to the earliest epochs of star formation, it will be possible to strengthen these constraints by another order of magnitude or more, giving us highly sensitive information about the earliest stages of nonlinear structure formation.

\section*{Author Contributions}

C.B., M.S. and A.C.J. proposed the idea for this project.
M.M. performed the $N$-body simulations, halo-finding, and halo mass function analysis with input from Y.W., S.B. computed the galaxy populations, and A.C.J. made the gravitational-wave predictions.
All authors contributed to every stage of the paper, including the writing.

\section*{Acknowledgments}
The authors would like to thank Joe Silk for useful discussions, and Ilya Mandel for helpful correspondence about \textsc{compas}.
M.S. thanks the University of Sydney for its hospitality during the early stages of this project.
A.C.J. was supported by the Science and Technology Facilities Council through the UKRI Quantum Technologies for Fundamental Physics Programme (Grant No. ST/T005904/1).
S.B. is supported by the UK Research and Innovation (UKRI) Future Leaders Fellowship (Grant No. MR/V023381/1).
M.S. is supported in part by the Science and Technology Facility Council (STFC), United Kingdom, under the Research Grant No. ST/P000258/1.
Y.W. is supported in part by the Australian Government through the Australian Research Council’s Future Fellowship (Project No. FT180100031).
The authors acknowledge the Sydney Informatics Hub and the use of the University of Sydney’s high-performance computing cluster, Artemis.
This work used the DiRAC@Durham facility managed by the Institute for Computational Cosmology on behalf of the STFC DiRAC HPC Facility (www.dirac.ac.uk).
The equipment was funded by BEIS capital funding via STFC Capital Grants No. ST/P002293/1, No. ST/R002371/1 and No. ST/S002502/1, Durham University and STFC Operations Grant No. ST/R000832/1.
DiRAC is part of the National e-Infrastructure.
This work was partly enabled by the UCL Cosmoparticle Initiative. 
Simulations in this paper made use of the \textsc{compas} rapid binary population synthesis code (version 02.21.00), which is freely available at Ref.~\cite{compas-github}.
This paper has an Einstein Telescope Document No. ET-0171A-22.

\bibliographystyle{apsrev4-1}
\bibliography{references}
\end{document}